\newcommand{\comment}[1]{} 

\newcommand{\rc}{r_{\rm c}}

\newcommand{\omegas}{\Omega_{\rm s}}
\newcommand{\omegak}{\Omega_{\rm k}}

\newcommand{\fim}{\varphi} 
\newcommand{\pd}[2]{\frac{\partial #1}{\partial #2}} 
\newcommand{\betat}{\tilde{\beta}}
\newcommand{\bt}{\tilde{b}_{\varphi}}

\documentclass{pazh2col}

\usepackage{graphicx}
\usepackage{physics}
\usepackage{amsmath}
\usepackage{ amssymb }

\begin{document}


\authors[Kuzin A. V.]{
  \nextauth[alv.kuzin@gmail.com]{A. V. Kuzin}{1}
}


\titles[Induced magnetic field in accretion disks]{Induced magnetic field in accretion disks around neutron stars}


\affiliations{
\nextaffil{Sternberg Astronomical Institute, Moscow, Russia}
}


\wideabstract{
There are X-ray pulsating sources that are explained by accretion from disks around neutron stars. Such disks deserve a detailed analysis. In particular, the dipole magnetic field of the central star may penetrate the disk, giving rise to an induced magnetic field inside the disk due to the frozen-in condition. The growth of the induced field can be limited by the turbulent diffusion. In the present work, I calculate the induced field in this case. The problem is reduced to the induction equation to which I have found an analytical solution describing  radial and vertical structures of the induced field. The radial structure is close to the earlier predicted dependence on the difference in angular velocities between the magnetosphere and disk, $b \propto \omegas - \omegak$, while the vertical structure of an induced field is close to the linear dependence on the altitude above the equator, $b \propto z$. The possibility of the existence of non-stationary quasi-periodic components of the induced field is discussed. 

\keywords{neutron stars, accretion disks, magnetic field}

\doi{... }
} 


\section{Introduction}
\label{introduction}
The problem of the structure of an accretion disk interacting with the magnetic field of a central star is closely related to X-ray sources with magnetized neutron stars (NSs). Modeling the disk, one needs to know the induced magnetic field inside and around the disk in order to properly account for the effects of the magnetic field. The situation when the NS magnetic field penetrates a disk is considered, the magnetic axis of the NS is inclined to the disk axis. Since the disk consists of ionized matter and the field lines are twisted by the movement of the disk, an induced field appears in the disk (Lai, 1999).

One of the first models of interaction between magnetic field and accretion disk was proposed by Gosh and Lamb (Gosh et al., 1977; Gosh and Lamb, 1979a, b) to explain observed variations of NSs spin rates. They calculated the exchange of torque between the disk and the star through the magnetic field. At the time, their theoretical results were consistent with observations. However, these works introduced the induced field in a simplified way. Wang (1987) has demonstrated an inconsistency of the model of Gosh and Lamb. This problem motivated other authors to build more realistic models of the induced field.

Campbell (1987) found the structure of induced magnetic field by solving the induction equation, using assumptions similar to the ones I use in this paper (seel also Campbell 1992). He proposed that the induced field is proportional to the difference in angular velocities between the magnetosphere and the disk: $b \propto (\Omega_s - \Omega(r))$. Wang (1995 and 1997 for the inclined rotator) examined several mechanisms limiting the growth of the induced magnetic field and found that turbulent diffusion could limit the magnetic field inflation. Wang also studied the case of reconnection of field lines above the disk in regions of large velocity differences (see also Lovelace et al., 1995).

The major points of the model presented below are as follows. In order to calculate the structure of accretion disk, one needs to know both the radial and vertical structure of the induced magnetic field. In this work, the radial and vertical components of the induced field are separated. The magnetic diffusion coefficient varies with the radial coordinate in a physically plausible manner. The equation for the induced magnetic field is derived from the induction equation, and boundary conditions are set for this equation. The focus of this work is on the stationary axially symmetric induced magnetic field, although other possibilities are also discussed.

\section{Model of disk and neutron star}
\label{sec:model}
The cylindrical stationary coordinate system $(r, \fim, z)$ is used with the center at the NS. The NS spins with frequency $\omegas$ and its spin axis is aligned with the disk axis; the disk equator lies in the $z = 0$ plane. The disk surface' coordinate is denoted as $z_0(r)$. The magnetic field of the NS is assumed to be dipolar with moment $\vb{\mu}$ and with the magnetic axis inclined to the spin axis at the angle $\chi$.

The accretion disk is geometrically thin, and its relative half-thickness $h_0 = z_0(r)/r \ll 1$ is a parameter of the model. Since $h_0$ is a weak function of a distance in the standard theory of accretion, I assume $h_0 = \textrm{const}$. The inner radius of the disk is denoted as $r_0$. Although the position of $r_0$ depends on the magnetic field (including the induced field), it is considered a free parameter in this work. For simplicity, $r_0$ demarcates the regions with matter corotating with the NS (magnetosphere) and matter in the disk. Instead of modeling a transition zone (as done, for example, in Kluźniak and Rappaport, 2007), a sharp transition from one zone to the other is assumed. The same sharp transition is assumed to occur at the surfaces of the disk. The matter inside the disk has Keplerian velocity, and deviations from which are likely unimportant for our problem (see Campbell, 1987). Outside of the disk, the matter corotates with the NS:

\begin{equation}\label{velocity}
    V = 
\begin{cases}
\displaystyle
\left(\frac{GM}{r}\right)^{1/2}, & \text{inside the disk}, \\
\displaystyle
\omegas r, & \text{in magnetosphere}.
\end{cases}
\end{equation}

The magnetic field is assumed to consist of dipolar components that have penetrated the disk and an additional toroidal field: 
\begin{equation}\label{field_gen}
    \vb{B} = \vb{B}_{\rm NS}^{\rm dip} + b_{\fim} \vb{e}_{\fim} = B_r \vb{e}_{r} + (B_{\fim 0} + b_{\fim})\vb{e}_{\fim} + B_z \vb{e}_{z},
\end{equation}
where $b_{\fim} = b_{\fim}(r, \fim, z)$ is the induced field.

I define the radius of corotation: $\rc = (GM / \omegas^2)^{1/3}$ and fastness parameter: $\omega = \omegas/\omegak(r_0) = (r_0 / \rc)^{3/2}$ as usual. All results in this work are given for the parameters typical for accreting millisecond pulsars: 
$M = 1.4 M_{\odot}$, $\mu = 10^{26}\textrm{~G~} \textrm{cm}^3$, $f = 200 \textrm{~Hz}$.
\section{Main equations}
We shall start with the magnetic field induction equation (e.g., Naso and Miller 2010, 2011):
\begin{equation}
        \pd{\bf{B}}{t} = \textrm{rot} \left([ \vb{\upsilon} \times \bf{B}] - \eta \textrm{ rot} \bf{B}\right).
    \label{general_induct}
\end{equation}

Magnetic diffusion coefficient $\eta$ is assumed to be of turbulent nature and proportional to the coefficient of kinematic viscosity: $\eta \propto \nu_{\rm T}$. The latter can be expressed from the alpha parameter (see, e.g., Lipunova 2018), $\nu_{\rm T} = 2\alpha z_0^2 \omegak/3\Pi_1$. $\Pi_1$ depends weakly on the radius, $\Pi_1 \approx 6-7$. Based on this formula, one may write: 
\begin{equation}\label{eta}
\eta = \omegak r^2 / C.
\end{equation}
If a coefficient of proportionality is introduced: $\eta = \epsilon \nu_{\rm T}$, then $C = 3\Pi_1/2\alpha\epsilon h_0^2$. I assume $C$ not to depend on radius, since both $h_0$ and $\Pi_1$ are weak functions of $r$. Possible relations between $\eta$ in the disk and magnetosphere are discussed below. The dipole magnetic field in cylindrical coordinates can be written as follows:  
\begin{multline}\label{dipole_field}
        \vb{B}^{\rm dip} =
        \frac{\mu}{(r^2 + z^2)^{5/2}} \times \\  \times [(2r^2 \sin{\chi} \cos{\varphi'} + 3rz \cos{\chi} - z^2 \sin{\chi} \cos{\varphi'}) \vb{e}_r +\\
        + (\sin{\chi}\sin{\varphi'} (r^2 + z^2))\vb{e}_{\varphi'} + \\
        + (-r^2\cos{\chi} + 3rz \sin{\chi} \cos{\varphi'} + 2z^2 \cos{\chi})\vb{e}_z].
\end{multline}
Here the angle $\fim'$ in the rotating reference system is related to the angle in the inertial reference frame $\fim$ as:  $\fim' = \fim - \Omega_s t$.

\subsection{Equation of field diffusion}

To estimate the strongest possible induced field inside the disk, I assume that the NS dipolar field penetrates the disk without screening (the assumption is discussed further). 

Since the field changes quasi-periodically and since the time $t$ enters all expressions in combination $(\fim - \omegas t)$, the time derivative may be replaced with: $\pd{}{t} = -\omegas \pd{}{\fim}$. Mind the following relations: $\textrm{rot}B^{\rm dip} = \vb{0}$; $\textrm{grad}\eta = (\eta/2r) \vb{e}_{r}$. The toroidal projection of the induction equation takes the form:

\begin{multline}\label{phi_comp}
    -\omegas \pd{B_{\fim}}{\fim} = \pd{(VB_z)}{z} + \pd{(VB_r)}{r} +\\
    +\eta \left(\Delta b_{\fim} - \frac{b_{\fim}}{2r^2} + \frac{1}{2r} \pd{b_{\fim}}{r}\right).
\end{multline}

The equation requires boundary conditions, but the expression in the last brackets in \eqref{phi_comp} is not suitable for setting them. It is convenient to make a substitution:
\begin{equation}
    \tilde{b}_{\fim} = \left(\frac{r}{\rc}\right)^{1/4} b_{\fim}    
\end{equation}
that eliminates the first $r$-derivatives in the brackets. Using velocity \eqref{velocity} and magnetic field \eqref{field_gen} with dipole field \eqref{dipole_field} (only terms $(z/r)^n, n \leq 1$, are considered), we obtain the induction equation \eqref{phi_comp} in the form:
\begin{multline}\label{eq_for_bphi}
    \left(\frac{\rc}{r}\right)^{1/4}\left(\Delta \bt - \frac{9}{16r^2} \bt + \frac{C}{r^2} \frac{\omegas}{\omegak}\pd{\bt}{\fim} \right)= \\
    =\frac{C\mu}{r^5} (4 + \frac{\omegas}{\omegak}) \sin{\chi} \cos{\fim} + \frac{9C\mu}{2r^6}z  \cos{\chi}  - \\
    - \frac{C\mu}{r^5} \frac{D_z}{\omegak}(-\cos{\chi} + 3z/r \cos{\fim}\sin{\chi})-\\
    - \frac{C\mu}{r_0^5}\frac{D_r}{\omegak}(2\sin{\chi}\cos{\fim}+3z/r \cos{\chi}).
\end{multline}
Coefficients $D_r$ and $D_z$ include delta-functions from the velocity derivatives:
\begin{equation}
    D_r = \omegas r_0 \frac{1 - \omega}{\omega} \delta(r - r_0),
\end{equation}
\begin{equation}
    D_z = r(\omegas - \omegak)(\delta(z-z_0) - \delta(z+z_0)).
\end{equation}

The angle $\fim = \omegas t + \fim'$ enters  equation \eqref{eq_for_bphi} exclusively as $\cos{\fim}$. Thus, the induced field consists of angle-independent component and possibly of components proportional to $\cos{\fim}$ and $\sin{\fim}$ only. Let us denote $b_{\fim} = b(r, z) +  b_1(r, z) \cos{\fim} + b_2(r, z) \sin{\fim}$. The first and second terms on the right-hand side of equation \eqref{eq_for_bphi} are sources for  $b_1$ and $b$. The third and fourth terms are responsible for the field discontinuity at the surfaces and inner radius of the disk.  These terms will be used for setting the boundary conditions. Inside the disk, for the axially symmetric component $\tilde{b} = \tilde{b}(r, z)$  the following equation may be obtained:
 \begin{equation}\label{eq_for_b_nonphi}
     \frac{1}{r} \pd{}{r}\left(r\pd{\tilde{b}}{r}\right) - \frac{9\tilde{b}}{16r^2} + \frac{\partial^2 \tilde{b}}{\partial z^2} = \frac{9C}{2}\frac{z}{r^6} \left(\frac{r}{\rc}\right)^{1/4} \mu \cos{\chi}.
 \end{equation}
 
The boundary condition at the surface of the disk will  be a condition of the second kind. To obtain it, I integrate \eqref{eq_for_bphi} over the small segment between two close points, one being in the disk and the other in the magnetosphere: $[z_0(1 - \varepsilon), z_0(1 + \varepsilon)]$, with $\varepsilon \ll 1$. Some axially-symmetric components in \eqref{eq_for_bphi} give zero upon integration. Let us rewrite \eqref{eq_for_bphi} leaving only the terms which yield non-zero terms upon integration:
\begin{equation}\label{eq_bound_cond_gen}
    \eta \left(\frac{\rc}{r}\right)^{1/4} \frac{\partial^2 \tilde{b}}{\partial z^2} = \frac{\mu\cos{\chi}}{r^3}(V_{\rm m} - V_{\rm d}) \delta(z-z_0) + ...
\end{equation}

We see that the discontinuity of the derivative $\pd{b}{z}$ at the surface of the disk is defined by the difference between disk and magnetosphere velocities. In order to obtain the condition for the derivative inside the disk, one needs to know how it relates to the derivative in the magnetosphere. There are two ways to establish the connection.

The first approach (Naso and Miller 2010, 2011; Rekowski et al. 2000) is based on the assumption that similar turbulent mechanisms are responsible for the magnetic diffusion both in disk and magnetosphere, but, since the density of matter is low and temperature is high above the disk (in the corona), $\eta_{\rm magn} \gg \eta_{\rm disk}$. If this is the case, we move $\eta$ to the r.h.s. of the equation  \eqref{eq_bound_cond_gen} and integrate over the small segment containing a surface of the disk. Using \eqref{eta}, we obtain:
\begin{equation}\label{bound_N_z}
    \pd{\tilde{b}}{z}\bigg|^{z = z_0(1 + \varepsilon)}_{z = z_0(1 - \varepsilon)} = -\frac{C \mu \cos{\chi}}{2} \left(1 - \frac{\omegas}{\omegak}\right) \frac{1}{r^4} \left(\frac{r}{\rc}\right)^{1/4}.
\end{equation}
Now the terms $\pd{\tilde{b}}{z}$ in the disk and magnetosphere should be connected somehow. Notice that, for a fixed $r$, the term $[\vb{v} \times  \vb{B}]$ in \eqref{general_induct} is of the same order under and above the disc surface\footnote{Only the poloidal field enters $[\vb{v} \times \vb{B}]$, and this poloidal field is  the continuous dipolar poloidal field.}. Hence, for the stationary field :
\begin{equation}\label{est}
    \eta_{\rm magn} \frac{\partial^2 \tilde{b}_{\rm magn}}{\partial z^2} \sim \eta_{\rm disk} \frac{\partial^2 \tilde{b}_{\rm disk}}{\partial z^2}.
\end{equation}
Assume that the induced field scale height in the disk is $\Delta z \sim z_0$. Let us denote the scale height for $b$ (or $\tilde{b}$) variation in the magnetosphere as $z_m$. Then, changing derivatives to finite differences in \eqref{est}, we get:
\begin{equation}
    \eta_{\rm magn} \frac{\tilde{b}_{\rm magn}}{z_{\rm m}^2} \sim \eta_{\rm disk} \frac{\tilde{b}_{\rm disk}}{z_{\rm d}^2}.
\end{equation}
The non-dipolar component of the field is expected to be of the same order below and above the surface, thereby:
\begin{equation}
    \pd{\tilde{b}_{\rm magn}}{z} \sim \frac{\tilde{b}_{\rm magn}}{z_{\rm m}} \sim \sqrt{\frac{\eta_{\rm d}}{\eta_{\rm magn}}} \pd{\tilde{b}_{\rm disk}}{z} \ll \pd{\tilde{b}_{\rm disk}}{z}.
\end{equation}
Thus, assuming turbulent diffusion in the magnetosphere, one may ignore the induced field derivative there. I follow this approach because the presence of corona seems very likely. From  \eqref{bound_N_z} we get:
\begin{equation}\label{bound_cond_1}
     \pd{\tilde{b}}{z}\bigg|_{z = z_0 - \varepsilon} = \frac{C \mu \cos{\chi}}{2} \left(1 - \frac{\omegas}{\omegak}\right) \frac{1}{r^4} \left(\frac{r}{\rc}\right)^{1/4}.
\end{equation}
There are certain flaws in this approach: if the corona corotates with the NS, then it is probably non-turbulent. It is unclear whether the magnetic diffusion coefficient in the magnetosphere is higher that the one on disk in this case.
The alternative approach to describe the magnetic diffusion coefficient suggests to assume the complete absence of matter in the magnetosphere, $\eta_{\rm magn} = 0$ (Campbell 1992). In that case, integrating \eqref{eq_bound_cond_gen}, we obtain:
\begin{equation}\label{bound_cond_2}
     \pd{\tilde{b}}{z}\bigg|_{z = z_0(1 - \varepsilon)} = C \mu \cos{\chi}\left(1 - \frac{\omegas}{\omegak}\right) \frac{1}{r^4} \left(\frac{r}{\rc}\right)^{1/4}.
\end{equation}
It is remarkable how little the result depends on the choice for setting the boundary conditions. 

The boundary condition at the lower surface of the disk may be set in the same fashion. At the inner radius of the disk, by analogy, one may write:
\begin{equation}\label{bound_r}
    \pd{\tilde{b}}{r}\Bigg|_{r = r_0} =
    -\frac{3C(1-\omega)}{2}\frac{1}{r_0^4} \left(\frac{r_0}{\rc}\right)^{1/4}\frac{z}{r_0} \mu \cos{\chi}.
\end{equation}

The number of boundary conditions is still not enough for the problem. There is a consensus that, if a disk exists, the magnetic field retains the dipole configuration only near the NS (Lovelace et al. 1995, Uzdenski et al. 2002). Far from the star, lines disconnect from the star due to reconnection, so there should be no induced field there. I set an additional boundary condition of the first kind at some outer radius $r_{\rm out}$. This radius cannot be determined self-consistently in our approach, making it an additional parameter. The outer radius in units of the corotation radius is denoted as $a$: $a = r_{\rm out}/\rc$.
The full problem for $\tilde{b}$ is:
\begin{equation}\label{for_b_nonphi}
\begin{cases}
\displaystyle
        \frac{1}{r} \pd{}{r}\left(r\pd{\tilde{b}}{r}\right) - \frac{9\tilde{b}}{16r^2} + \frac{\partial^2 \tilde{b}}{\partial z^2} = \frac{9C}{2} \frac{z}{r^6} \left(\frac{r}{\rc}\right)^{1/4} \mu \cos{\chi},\\
\displaystyle
        \pd{\tilde{b}}{z}\Bigg|_{z = \pm z_0} = \frac{C}{2} \frac{1}{r^4} \left(1 - \frac{\omegas}{\omegak}\right) \left(\frac{r}{\rc}\right)^{1/4}\mu \cos{\chi},\\
\displaystyle
        \pd{\tilde{b}}{r}\Bigg|_{r = r_0} = -\frac{3C}{2r_0^5} (1-\omega)\left(\frac{r_0}{\rc}\right)^{1/4} z\mu \cos{\chi},\\
\displaystyle
        \tilde{b}(r = r_{\rm out}) = 0.\\
\end{cases}
\end{equation}

Note that $b$ is proportional to the dipole moment of the NS and cosine of the inclination angle: $b \propto \mu \cos{\chi}$.
It is convenient to normalize the field $\tilde{b}$  by the value $C \mu \cos{\chi}$, which I will denote as $\mu_c$. It is easier to solve the problem with trivial boundary conditions at the surface, so I rewrite \eqref{for_b_nonphi} for a new function  $\beta$:
\begin{multline}\label{substitution}
        \tilde{b}(r, z) = \\
        = \frac{\mu_c}{2r^4}\left(1 - \frac{\omegas}{\omegak}\right)\left(\frac{r}{\rc}\right)^{1/4} z + \frac{\mu_c}{\rc^3}\betat(\rho, \zeta) \equiv\\
        \equiv \tilde{b}_0(r, z) + \frac{\mu_c}{\rc^3}\betat(\rho, \zeta).
\end{multline}
Here dimensionless quantities are introduced: $\rho = r/\rc$, $\zeta = z/z_0 = z/(rh_0)$.
Before finding $\tilde{\beta}$, let us note that $\tilde{b}_0$ component in \eqref{substitution} is determined by the non-zero boundary conditions at the surface of the disk and is proportional to the gap in velocity of matter at the surface.  Moving from $\tilde{b}_0$  back to $b_0$  and setting the vertical coordinate to be coordinate of the disk surface: $z = z_0(r) = h_0 r$,  we get $b_0^{\rm surf} = (\mu_c h_0/2)  (1 - \omegas / \omegak) / r^3$. Up to a numerical factor, this is the expression given by Wang for the induced field at the surface of the disk in case of turbulent diffusion being the factor for limiting the field growth. This expression is frequently used as an analytic estimation for the induced field at the surface of the disk (see, e.g., Klu$\acute{z}$niak and Rappaport 2007). Thus,  $b_0(r, z)$ describes the vertical distribution of the field found by Wang at the surface.

The other component of the field, $\betat$, is determined by the boundary conditions at the inner radius of the disk and by the right-hand side of equation \eqref{for_b_nonphi} and was not considered by Wang. For the convenience of comparing our results, the coefficient $C$ is set so that the formula for $b_0^{\rm surf}$ coincides with the expression used by Klu$\acute{z}$inak and Rappaport, that is, $C = 2/h_0$.

We obtain the following problem for $\betat(r, z)$:
\comment{
\begin{equation}\label{eq_for_y}
\begin{cases}
\displaystyle
        \frac{1}{r} \pd{}{r}\left(r\pd{\betat}{r}\right) - \frac{9\betat}{16r^2} + \frac{\partial^2 \betat}{\partial z^2} = F(r, z),\\
\displaystyle
        \pd{\betat}{z}\Bigg|_{z = z_0(r)} =  \pd{\betat}{z}\Bigg|_{z = -z_0(r)} = 0,\\
\displaystyle
        \pd{\betat}{r}\Bigg|_{r = r_0} = \frac{3\mu_c}{8r_0^5}(1+\omega)z \omega^{1/6},\\
\displaystyle
        \betat(r = r_{\rm out}) = -\frac{\mu_c}{2} \frac{z}{\rc^4} \frac{1-a^{3/2}}{a^4} a^{1/4},\\
\displaystyle
    F(r, z) = \frac{9\mu_c}{4} \frac{z}{r^6} \left(\frac{\omegas}{\omegak}-1\right)\left(\frac{r}{\rc}\right)^{1/4},     
\end{cases}
\end{equation}
}
\begin{equation}\label{eq_for_y_b}
\begin{cases}
\displaystyle
        \frac{1}{\rho} \pd{}{\rho}\left(\rho\pd{\betat}{\rho}\right) - \frac{9\betat}{16\rho^2} + \frac{1}{h_0^2 \rho^2}\frac{\partial^2 \betat}{\partial \zeta^2} = f(\rho, \zeta),\\ 
        \displaystyle
        \pd{\betat}{\zeta}\Bigg|_{\zeta = 1} = 0,  \betat(\zeta = 0) = 0, \\
        \displaystyle
        \pd{\betat}{\rho}\Bigg|_{\rho = \omega^{2/3}} = \frac{3}{4} \frac{1 + \omega}{\omega^{8/3}} \omega^{1/6} \zeta,\\
        \displaystyle
        \beta(\rho = a) =  \frac{a^{3/2} - 1}{a^3} a^{1/4} \zeta,\\
        \displaystyle
        f(\rho, \zeta) = \frac{9}{2} \frac{\zeta}{\rho^5} \left(\rho^{3/2} - 1\right) \rho^{1/4}.
\end{cases}
\end{equation}
which is solved in the next section. Notice that in case of a very thin disk: $h_0 \rightarrow 0$ and $\beta(r, z) \rightarrow 0$. This means that the deviation of the induced field $b$ from the ``Wang's field'' $b_0$ is absent in case of the very small relative thickness and increases as $h_0$ increases.

\section{The solution for the induced field}\label{sec:ind_field}

\begin{figure}
 \includegraphics[width=\columnwidth]{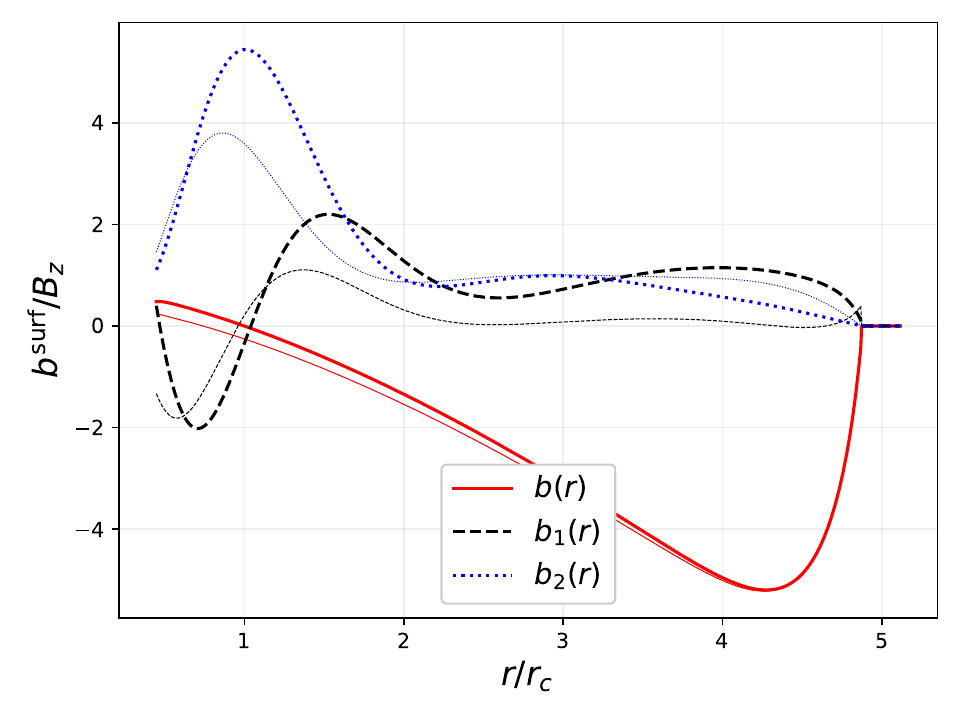}
 \caption{Comparison of the induced field, calculated in cylindrical coordinate system, with the field numerically computed in the spherical coordinate system (thin lines of the same color). Relative half-thickness of the disk  $h_0 = 0.1$, fastness parameter $\omega = 0.3$. The field  $b$ and Fourier components  $b_1, b_2$ (see text) at the disk surface are shown.}
 \label{fig:appendix_b}
\end{figure}

Solving \eqref{eq_for_y_b}, we  assume that $\frac{\partial^2 \betat}{\partial z^2} \approx \frac{1}{h_0^2 r^2} \frac{\partial^2 \betat}{\partial \zeta^2}$. This is mathematically incorrect, but the full expression would be so complicated that the solution with the method of separation of variables would not be feasible. The reason for this complication is that the region in which the equations are to be solved is a truncated trapeze. At the surfaces of this trapeze, the boundary conditions are set in form of $\pd{\betat}{z} = ...$, instead of a canonical form $\pd{\betat}{\vb{n}} = ...$. Replacement of $\pd{\betat}{z}$ with $\frac{1}{h_0r}\pd{\betat}{\zeta}$ in the equations is  equivalent to replacing the partial derivatives along the $z$-coordinate in the boundary conditions with the derivatives along the normal $\vb{n}$. Since the disk opening angle is about $2h_0 \ll 1$, one can expect deviation of our solution from the exact one to be small.  Such situation is absent in the spherical coordinate system (SphCS), and it is actually easier to solve the induction equation in CphCS than in the cylindrical coordinate system (CyCS). Nonetheless, with the assumption discussed, the  'vertical' eigenfunctions in CyCS are simple $\sin{(\mu_n x)}$ (see below), while the eigenfunctions in the CphCS are more complicated functions.  Induced magnetic field calculated in both CyCS (with the simplifications described above) and in CphCS is plotted in Figure \ref{fig:appendix_b}. Although the deviations are noticeable, they are insignificant in the context of disk structure modelling.

It is clear that $\betat$ is an odd function of $\zeta$. Consider the problem on segment $[0, 1]$ instead of $[-1, 1]$. One needs to replace the lower surface boundary condition by the condition at the equator: $\beta(\zeta = 0) = 0$. Let us search the solution in the following form: 
\begin{equation}\label{expansion}
    \betat(\rho, \zeta) =   \sum_{n=0}^{\infty} B_n(\rho) \sin{(\mu_n \zeta)},
\end{equation}
where $\mu_n = \pi(n + 1/2)$. For  $B_n(\rho)$ coefficients we obtain one-dimensional boundary problems. Using linearity of equations, one may split $B$ into the sum of two components\footnote{This step is necessary for the analytical solution of \eqref{eq_for_y_b}, but in reality it is faster and easier to solve  boundary problems for $B_n(r)$ numerically (for a suitable number of  $n$). The analytical solution is useful at least for checking the code.}: $B(\rho) = u(\rho) + v(\rho)$. The problem for $u$ includes zero boundary conditions and non-zero source, while the problem for $v$ lacks the source, but includes the non-trivial boundary conditions:
\begin{equation}
\begin{cases}
    \displaystyle
    \rho(\rho u_n')' - M_n^2 u_n = \rho^2 f_n(\rho),\\ 
        \displaystyle
    u_n'(\omega^{2/3}) = 0, u_n(a) = 0,\\
    \displaystyle
    f_n(\rho) = 9 \frac{(-1)^n}{\mu_n^2} \frac{\rho^{1/4}}{\rho^5}\left(\rho^{3/2} - 1\right).
\end{cases}
\end{equation}
\begin{equation}
\begin{cases}
    \displaystyle
   \rho(\rho v_n')' - M_n^2 u_n = 0,\\
        \displaystyle
    v_n'(\omega^{2/3}) = 2 \frac{3}{4} \frac{(-1)^n}{\mu_n^2}\frac{1 + \omega}{\omega^{8/3}} \omega^{1/6},\\
        \displaystyle
    v_n(a) = 2 \frac{(-1)^n}{\mu_n^2} \frac{a^{3/2} - 1}{a^3} a^{1/4}.\\
    \displaystyle
\end{cases}
\end{equation}
Here I denote $M_n^2 = 9/16 + \mu_n^2 / h_0^2$. 
The problem for $v$ has the solution:
\begin{equation}\label{vn}
    v_n(\rho) = A_n \rho^{M_n} + C_n \rho^{-M_n},
\end{equation}
where $A, C$ may be found easily from boundary conditions. 

The problem for $u$ may be solved in the following way. First, we solve the Sturm-Liouville problem with corresponding boundary conditions:
\begin{equation}\label{eq_for_u_short}
\begin{cases}
\displaystyle
        \frac{1}{\rho} (\rho\tilde{y}_{mn}')' - \frac{M_n^2}{\rho^2}\tilde{y}_{mn} + \lambda_{mn} \tilde{y}_{mn} = 0,\\
        + \text{zero B.C.}
\end{cases}
\end{equation}

Eigenfunctions of \eqref{eq_for_u_short} are combinations of Bessel functions of the first and second kind:
\begin{equation}
    \tilde{y}_{mn} = J_{M_n}(\rho\sqrt{\lambda_{mn}}) - N_{M_n}(\rho\sqrt{\lambda_{mn}})  \frac{J_{M_n}(a\sqrt{\lambda_{mn}})}{N_{M_n}(a\sqrt{\lambda_{mn}})},
\end{equation}
while the eigenvalues are roots of the equation:
\begin{equation}
    \frac{J'_{M_n}(\omega^{2/3}\sqrt{\lambda_{mn}}) }{J_{M_n}(a\sqrt{\lambda_{mn}})} = \frac{ N'_{M_n}(\omega^{2/3}\sqrt{\lambda_{mn}})}{N_{M_n}(a\sqrt{\lambda_{mn}})} .
\end{equation}

Then 
\begin{equation}\label{un}
    u_n(\rho) = -\sum_{m} \frac{f_{mn}}{\lambda_{mn}} \tilde{y}_{mn}(\rho),
\end{equation}
where
\begin{equation}
    f_{mn} = \frac{1}{||\tilde{y}_{mn}||^2} \int_{\omega^{2/3}}^{a} f(\rho)\tilde{y}_{mn}(\rho) \rho \text{d} \rho,
\end{equation}
\begin{equation}
    ||\tilde{y}_{mn}||^2 = \int_{\omega^{2/3}}^a \rho \tilde{y}_{mn}^2 \text{d}\rho.
\end{equation}

\begin{figure}
\centering
\includegraphics[width=\columnwidth]{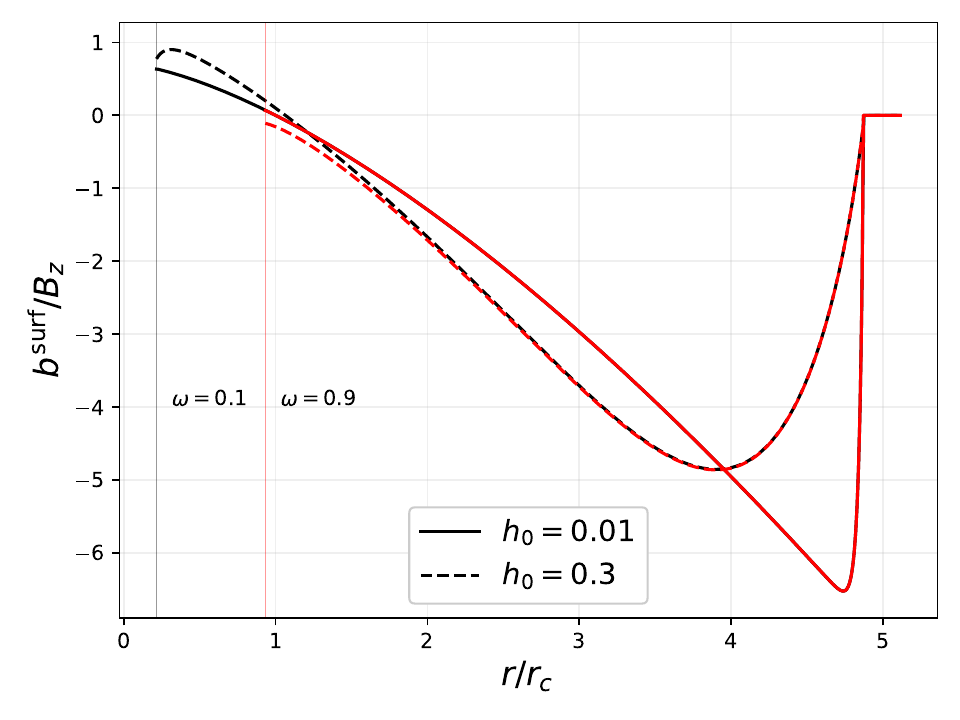}
\caption{The dependence of the induced field at the disk surface $b^{\rm surf}$, normalized by the vertical field $B_z$, on the distance to the NS in units of corotation radius. The field is calculated for two different fastness parameters ($\omega = 0.1$ and $\omega = 0.9$ correspond to red and black, inner radii are denoted by vertical lines). For each $\omega$, two relative half-thicknesses $h_0$ were adopted, see legend.}
\label{fig:rad}
\end{figure}

Finally, the analytical solution for the induced field can be found using \eqref{substitution}, \eqref{expansion}, \eqref{vn}, \eqref{un}. 
Radial distribution of the induced field at the surface of the disks shown in Figure \ref{fig:rad} for different fastness parameters $\omega$ (different disk inner radii $r_0 = \omega^{2/3} \rc$) and relative half-thicknesses $h_0$. For illustration purposes, the outer radius  $r_{\rm out}$ in the problem is set to be equal to the light cylinder radius. The reason is that the light cylinder radius is the maximal distance from the NS where a corotating magnetosphere can exist. It can be seen that, for a given thickness of the disk, the solutions for different $\omega$ are self-similar far from the inner edge. Thus, the position of $r_0$ affects the induced field only near the inner radius.

\begin{figure}
\centering
\includegraphics[width=\columnwidth]{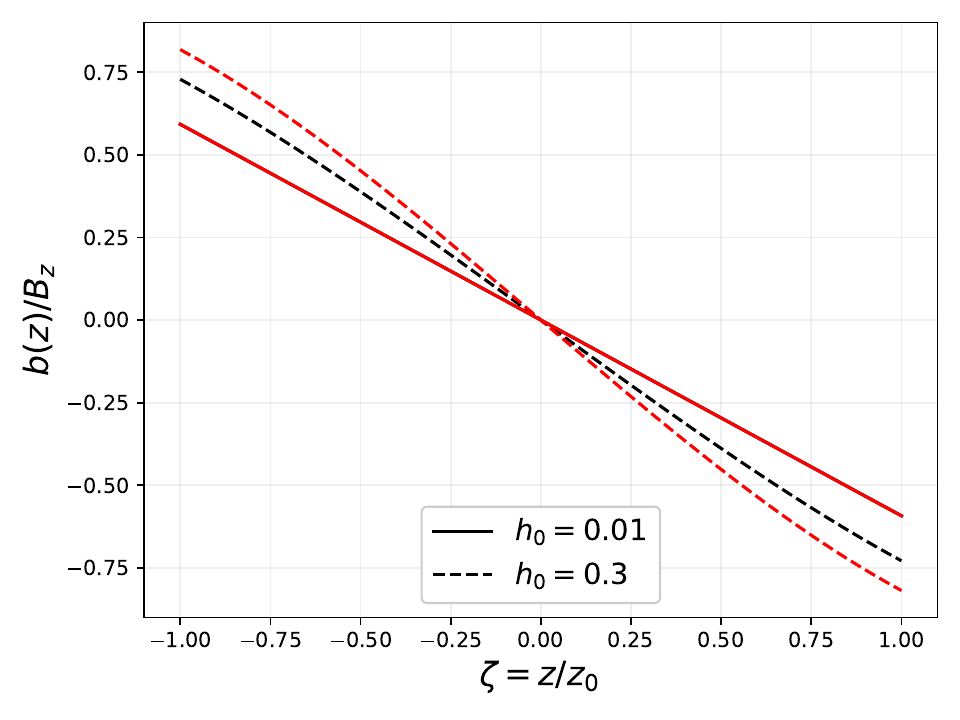}
\caption{Vertical distribution of the induced field inside the disk $b(z)$, normalized by the vertical dipole field $B_z$, at $r = 1.5 \rc$ for the same parameters  $\omega$ and $h_0$ as in Fig. \ref{fig:rad}. For the case of a thin disk, $h_0 = 0.01$, lines corresponding to $\omega = 0.1$ and $0.9$ merge.}
\label{fig:vert}
\end{figure}

Figure \ref{fig:vert} shows the vertical structure of the induced field for the same set of parameters. When the half-thickness is small,  the vertical profiles of the field closely follow linear proportionality: $b(\zeta) \propto \zeta$. For larger half-thickness, the dependence become more complicated, but still it can be described by linear function. This is due to the contribution of the ``Wang's component'' $b_0$ in the total induced field (see eq.  \eqref{substitution}), which is always a linear function of $z$. If $h_0 \ll 1$, then $\beta$ is also linear, which can be seen from the dimensionless equation \eqref{eq_for_y_b} for $\betat$.  

\section{Discussion}
Assuming that the growth of the induced field is limited by turbulent diffusion, three parameters are required to calculate it: the fastness parameter $\omega$, relative half-thickness $h_0$, and outer radius $a$ of the zone where the NS field remains dipole, in units of the corotation radius. This means that $\omega$ and $h_0$ should be calculated self-consistently when modeling the accretion disk using the found solution.

\subsection{Axisymmetric component}
Campbell (1987) assumed the same mechanism of magnetic field dissipation and found an axisymmetric stationary induced field in the form of an integral with a complicated kernel. The field was calculated at the half-line $[r_0, +\infty)$, while the vertical structure was obtained implicitly\footnote{The boundary conditions at the surfaces of the disk were not explicitly set by the author.}.

In the current work, the induced field $b$ was decomposed into a sum of $b_0 \propto (\omegas - \omegak) z$ and the modification $\beta$. The vertical coordinate enters the series for $\beta$ in a simple way. It is usually more convenient to analyze equations instead of two-dimensional integrals. Further, Campbell (1987) adopted two models for the magnetic diffusion coefficient: $\eta \sim \textrm{const}$ and $\eta \propto r^2$, while  $\eta \propto r^{1/2}$ is a better approximation for a disk with constant half-thickness if the proportionality between magnetic diffusion and turbulent kinematic viscosity coefficient is adopted.

One can account for the reconnection of the field lines by changing the position of $r_{\rm out}$. The process of reconnection is thought   to limit the induced field so that $|b/B_z| \lesssim 1$ (Wang 1995, Lovelace 1995). This condition may be met in our model by decreasing $a$. For relative half-thickness $h = 0.01$, the requirement $|b/B_z| < 1$ is met if $a \lesssim 1.9$. If $h = 0.3$, then one should set $a \lesssim 2.2$. We come to the conclusion that the zone where the NS field lines are connected to the disk should be narrow (see Matt and Pudritz 2005).
\subsection{Possible azimuthal-dependent contributions}
It is unclear to what extent the $\fim$-dependent components of the dipolar field penetrate the disk. In case they do not penetrate the disk at all, the induced field is given by $b(r, z)$. Lai (1999) attempted to build a model of the induced field in this scenario (the "Hybrid model").

On the other hand, if these components penetrate the disk, the approach from this work allows us to calculate non-asymmetric components of the induced field. Equation \eqref{eq_for_bphi} was derived, assuming complete penetration of the NS dipole field into the disk. According to it, the full induced field $b_{\fim}$ consists of $b$ and components proportional to sine and cosine of the azimuthal angle. Then $b_{\fim} = b(r, z) + b_1(r, z) \cos{\fim} + b_2(r, z) \sin{\fim}$. The equations for the Fourier components $b_{1, 2}$ can be derived in the same way as for $b$. Numerical solutions of these equations are displayed in Figure \ref{fig:appendix_b} (see also discussion at the beginning of Section \ref{sec:ind_field}). The components $b_{1, 2}$ are not small and would make a significant contribution to the magnetic field pressure if they existed.

One may suggest that the $\fim$-dependent component of the NS field only partially penetrates the disk. The skin depth can be estimated using the magnetic diffusion coefficient used in this work, however, this problem is beyond the scope of the current paper. Therefore, the Fourier coefficients $b_{1, 2}$ shown in Figure \ref{fig:appendix_b} are upper limits for the possible non-stationary components of the induced field. Since $b_{1, 2} \propto \mu \sin{\chi}$, they are absent in the case of alignment of the NS spin and magnetic axes.


\section{Conclusions}
\label{sect:concl}
In this work, I have focused on modeling the induced magnetic field in the accretion disk around a magnetized star with a tilted magnetic axis. By assuming that the central star's dipole magnetic field penetrates the disk, I derived a partial differential equation for the induced field. The solution for the axisymmetric component of the field was obtained using the method of separation of variables, allowing for both radial and vertical structure to be determined. The possible existence of non-axisymmetric components of the induced field was also discussed, with upper limits for their magnitude being presented.


The research was supported by RSF (No. 21-12-00141). Author thanks G. V. Lipunova for the productive discussion of the manuscript.



\begin{references}
\comment{
\reference
Гнедandн, Сюняев
(Yu.N. Gnedin and R.A. Sunyaev),
\aap\ \textbf{36}, 379 (1974).

\reference
Доandч (A.J. Deutsch),
Ann.\ d'Astrophys.\ \textbf{18}, 1 (1955).

\reference
andваненко Д.Д., Курдгелаandдзе Д.Ф.,
Астрофandзandка \textbf{1}, 479 (1965).

\reference
Манчестер Р., Теandлор Дж.,
 \textit{Пульсары} (М.: Мandр, 1980).

\reference
Трюмпер and др. (J. Tr\"umper, W. Pietsch, C. Reppin, W. Voges,
R. Staubert, and E. Kendziorra),
\apj\ \textbf{219}, L105 (1978).

\reference
Фортов В.Е.,
УФН \textbf{179}, 653 (2009).
}
\reference
Y.-M. Wang, 
\aap\ \textbf{183}, 257 (1987).

\reference
Y.-M. Wang, 
\apj\ \textbf{449}, 153 (1995).

\reference
Y.-M. Wang, 
\apj\ \textbf{475}, 135 (1997).

\reference
P. Ghosh, F. K. Lamb, C. J. Pethick, \apj\ \textbf{217}, 578 (1977).


\reference
P. Ghosh, F. K. Lamb, \apj\ \textbf{232}, 259 (1979).

\reference
P. Ghosh, F. K. Lamb, \apj\ \textbf{234}, 296 (1979).

\reference
W. Klu$\acute{z}$niak, S. Rappaport, \apj\ \textbf{671}, 1990 (2007).

\reference
C. G. Campbell, \mnras\ \textbf{229}, 405 (1987).

\reference
C. G. Campbell, Geophys. Astro. Fluid., \textbf{63}, 179 (1992).
\reference
R. V. E. Lovelace,  M. M. Romanova,  G. S. Bisnovatyi-Kogan, \mnras\ \textbf{275}, 244 (1995).

\reference
D. Lai, \apj\ \textbf{524}, 1030 (1999).

\reference
Lipunova G., Malanchev K., Shakura N. in Shakura N. ed., Astrophys. Space Sc. L. \textbf{454} (2018).

\reference
S. Matt, R. E. Pudritz, \apj\ \textbf{632}, 135 (2005).


\reference
L. Naso, J. C. Miller, \aap\ \textbf{521}, 31 (2010).

\reference
L. Naso, J. C. Miller, \aap\ \textbf{531}, 163 (2011).

\reference
M. V. Rekowski, G. R\"{u}diger, D. Elstner), \aap\, \textbf{353}, 813 (2000).

\reference
D. A. Uzdensky, A. K\"{o}nigl, C. Litwin, \apj\, \textbf{565}, 1191 (2002).

\reference
D. A. Uzdensky, A. K\"{o}nigl, C. Litwin, \apj\, \textbf{565}, 1205 (2002).

\end{references}
\end{document}